\documentclass[preprint,showpacs,amsmath,amssymb,superscriptaddress,showkeys,prc,nofootinbib]{revtex4}

\usepackage{graphicx}
\usepackage{dcolumn}
\usepackage{bm}
\usepackage{amsmath,amssymb}

\begin{document}


\title{Skewness of the elliptic flow distribution in $\sqrt{s_{_{\mathrm{NN}}}}$ = 5.02~TeV PbPb collisions from HYDJET++ model}

\author{P. Cirkovic}
\affiliation{Vin\v{c}a Institute of Nuclear Sciences, University of Belgrade, Mike Petrovi\'{c}a Alasa 12-14, 11351 Vinca, Belgrade, Serbia}

\author{J. Milosevic}
\email[Corresponding author:]{Jovan.Milosevic@cern.ch}
\affiliation{Vin\v{c}a Institute of Nuclear Sciences, University of Belgrade, Mike Petrovi\'{c}a Alasa 12-14, 11351 Vinca, Belgrade, Serbia}
\affiliation{University of Oslo, Department of Physics, Oslo, Norway}

\author{L. Nadderd}
\affiliation{Vin\v{c}a Institute of Nuclear Sciences, University of Belgrade, Mike Petrovi\'{c}a Alasa 12-14, 11351 Vinca, Belgrade, Serbia} 

\author{F. Wang}
\affiliation{College of Science, Huzhou University, Huzhou, 313000, P. R. China}
\affiliation{Department of physics and astronomy, Purdue University, Indiana 47907, USA}

\author{X. Zhu}
\affiliation{College of Science, Huzhou University, Huzhou, 313000, P. R. China}

\date{\today}

\begin{abstract}
The elliptic flow ($v_{2}$) event-by-event fluctuations in PbPb collisions at 5.02~TeV are analyzed within the HYDJET++ model. Using the multi-particle, so called Q-cumulant method, $v_{2}\{2\}$, $v_{2}\{4\}$, $v_{2}\{6\}$ and $v_{2}\{8\}$ are calculated and used to study their ratios and to construct skewness ($\gamma^{exp}_{1}$) as a measure of the asymmetry of the elliptic flow distribution. Additionally, in order to check if there is a hydrodynamics nature in the elliptic collectivity generated by the HYDJET++ model, the ratio of $v_{2}\{6\} - v_{2}\{8\}$ and $v_{2}\{4\} - v_{2}\{6\}$ distribution is calculated. The analysis is performed as a function of the collision centrality. In order to check the HYDJET++ model responses, the results of this analysis are compared to the corresponding experimental measurements by CMS. A good qualitative and rather good quantitative agreement is found.
\end{abstract}



\maketitle

\section{Introduction}

\label{intro}
In ultra-relativistic nucleus-nucleus collisions sufficiently high energy densities have been achieved that a new state of matter, the Quark-Gluon-Plasma (QGP) has been created. The QGP created in these collisions exhibits a collective expansion which could be described by relativistic hydrodynamic flows. The collectivity in the QGP has been studied in experiments at the Relativistic Heavy Ion Collider (RHIC)~\cite{Back:2002gz,Ackermann:2000tr,Adcox:2002ms} and the Large Hadron Collider (LHC)~\cite{Aamodt:2010pa,ALICE:2011ab,Abelev:2014pua,Adam:2016izf,ATLAS:2011ah,ATLAS:2012at,Aad:2013xma,Chatrchyan:2012wg,Chatrchyan:2012ta,Chatrchyan:2013kba,CMS:2013bza,Khachatryan:2015oea}. The geometry of the overlap interaction zone in nucleus-nucleus collisions is anisotropic. This anisotropy is converted into momentum space by the hydrodynamic expansion. The momentum anisotropy can be characterized by a Fourier expansion of the emitted hadron yield distribution in azimuthal angle, $\phi$,~\cite{Ollitrault:1993ba,Voloshin:1994mz,Poskanzer:1998yz}
\begin{equation}
\label{F1}
\frac{dN}{d\phi} \propto 1+2\sum_{n}v_{n}\cos[n(\phi - \Phi_{n})],
\end{equation}
where Fourier coefficients, $v_{n}$, represent magnitude of the azimuthal anisotropy measured with respect to the corresponding flow symmetry plane angle, $\Phi_{n}$. The flow symmetry plane is determined by the geometry of the participant nucleons and can be reconstructed from the emitted particles themselves. Because of fluctuations in the initial spatial geometry, all orders of Fourier harmonics are present. The second order Fourier coefficient, $v_{2}$, is called elliptic flow, while the angle $\Phi_{2}$ corresponds to the flow symmetry plane which is determined by the beam direction and the shorter axis of the roughly lenticular shape of the nuclear overlap region.

Another experimental method to determine the $v_{n}$ coefficients is multi-particle cumulant analysis which uses the Q-cumulant method~\cite{Bilandzic:2010jr}. The multi-particle cumulant technique has the advantage of suppressing short-range correlations arising from jets and resonance decays and revealing the collective nature of the observed azimuthal correlations. The two-, four-, six-, and eight-particle azimuthal correlations are calculated as:
\begin{eqnarray}
  \label{FExp}
  \left.\begin{aligned}
\langle\langle 2 \rangle\rangle &= \langle\langle e^{in(\phi_{1}-\phi_{2})} \rangle\rangle, \\
\langle\langle 4 \rangle\rangle &= \langle\langle e^{in(\phi_{1}+\phi_{2}-\phi_{3}-\phi_{4})} \rangle\rangle, \\
\langle\langle 6 \rangle\rangle &= \langle\langle e^{in(\phi_{1}+\phi_{2}+\phi_{3}-\phi_{4}-\phi_{5}-\phi_{6})} \rangle\rangle, \\
\langle\langle 8 \rangle\rangle &= \langle\langle e^{in(\phi_{1}+\phi_{2}+\phi_{3}+\phi_{4}-\phi_{5}-\phi_{6}-\phi_{7}-\phi_{8})} \rangle\rangle
\end{aligned}\right.
\end{eqnarray}
where $\langle\langle...\rangle\rangle$ denotes averaging over all particle multiplets and over all events from a given centrality\footnote{The centrality in heavy ion collisions is defined as a fraction of the total inelastic nucleus-nucleus cross section, with 0\% denoting the most central collisions.} class, $n$ is harmonic order and $\phi_{i}$ ($i = 1,...,8$) are the azimuthal angles of particles from a given particle multiplet. The corresponding multiparticle cumulants $c_{n}\{2k\}$ ($k = 1,...,4$) are then given as ~\cite{Bilandzic:2010jr}:
\begin{eqnarray}
  \label{Cumul}
    \left.\begin{aligned}
      c_{n}\{2\} &= \langle\langle 2 \rangle\rangle, \\
      c_{n}\{4\} &= \langle\langle 4 \rangle\rangle - 2 \langle\langle 2 \rangle\rangle^{2}, \\
      c_{n}\{6\} &= \langle\langle 6 \rangle\rangle - 9 \langle\langle 4 \rangle\rangle \langle\langle 2 \rangle\rangle + 12 \langle\langle 2 \rangle\rangle^{3}, \\
      c_{n}\{8\} &= \langle\langle 8 \rangle\rangle - 16 \langle\langle 2 \rangle\rangle \langle\langle 6 \rangle\rangle - 18 \langle\langle 4 \rangle\rangle^{2} \\
      &+ 144 \langle\langle 4 \rangle\rangle \langle\langle 2 \rangle\rangle^{2} - 144 \langle\langle 2 \rangle\rangle^{4}.
      \end{aligned}\right.
\end{eqnarray}
Finally, the Fourier coefficients $v_{n}$ are connected to the above defined multi-particle cumulants through the following relations
\begin{eqnarray}
  \label{FCoeff}
    \left.\begin{aligned}
      v_{n}\{2\} &= \sqrt{c_{n}\{2\}}, \\
      v_{n}\{4\} &= \sqrt[4]{-c_{n}\{4\}}, \\
      v_{n}\{6\} &= \sqrt[6]{\frac{1}{4}c_{n}\{6\}}, \\
      v_{n}\{8\} &= \sqrt[8]{-\frac{1}{33}c_{n}\{8\}}.
      \end{aligned}\right.
\end{eqnarray}

The unitless standardized skewness, $\gamma^{exp}_{1}$, of the event-by-event elliptic flow magnitude distribution is a measure of the asymmetry about its mean. This standardized skewness can be estimated using the cumulant elliptic flow harmonics defined as in Ref.\cite{Giacalone:2016eyu}
\begin{equation}
\label{Skew}
\gamma^{exp}_{1} = -6\sqrt{2}v_{2}\{4\}^{2}\frac{v_{2}\{4\}-v_{2}\{6\}}{(v_{2}\{2\}^{2}-v_{2}\{4\}^{2})^{3/2}}.
\end{equation}

In the case where the event-by-event elliptic flow magnitude fluctuations stem from an isotropic Gaussian transverse initial-state energy density profile, the skewness, $\gamma^{exp}_{1}$ becomes equal to zero. But, non-Gaussian fluctuations in the initial-state energy density profile could be present~\cite{Giacalone:2016eyu}, and as a consequence will produce differences in the higher order cumulant $v_{2}\{2k\}$ ($k\geq 2$) coefficients. As Eq.~(\ref{Skew}) is an approximation of the standardised skewness $\gamma_{1}$ defined in~\cite{Giacalone:2016eyu}, it is possible to test its validity through the universal equality given in~\cite{Giacalone:2016eyu}
\begin{equation}
\label{UnEq}
v_{2}\{6\}-v_{2}\{8\} = \frac{1}{11}(v_{2}\{4\}-v_{2}\{6\}).
\end{equation}

In this paper, we study the skewness of the elliptic flow distribution using the HYDJET++ model. The basic features of HYDJET++ model~\cite{Lokhtin:2008xi} are described in Sect.~2. Using HYDJET++ model, approximately 60M PbPb collisions at $\sqrt{s_{_{\mathrm{NN}}}}$ = 5.02~TeV are simulated and analyzed. The obtained results together with the corresponding experimental results~\cite{Sirunyan:2017fts,Acharya:2018lmh} and discussions are given in Sect.~3. The results are presented over a wide range of centralities going from central (5--10\% centrality) up to rather peripheral (55--60\% centrality) PbPb collisions. The analyzed $p_{T}$ interval is restricted to 0.3~$\le p_{T} \le $~3~GeV/c range where hydrodynamics dominates, while $\eta$ range covers (-1.0, 1.0) region. A summary is given in Sect.~4.

\section{HYDJET++}
\label{HYDJET}
In the Monte Carlo HYDJET++ model relativistic heavy ion collisions are simulated. The HYDJET++ model consists of two components which simulate soft and hard processes. The hydrodynamical evolution of the system is provided by the soft part of the model, while the hard part describes multiparton fragmentation within the formed medium. Within the hard part, jet quenching effects are also taken into account. The minimal transverse momentum $p^{min}_{T}$ of hard scattering of an incoming parton regulates whether it contributes to the soft or to the hard part. The partons which are produced with $p_{T} < p^{min}_{T}$, or which are quenched below $p^{min}_{T}$ do not contribute to the hard part. The hard part of the HYDJET++ model consists of PYTHIA~\cite{Sjostrand:2006za} and PYQUEN~\cite{Lokhtin:2005px} models. These models simulate initial parton-parton collisions, radiative energy loss of partons and parton hadronization. The soft part of the HYDJET++ model regulates the magnitude of the elliptic flow by corresponding spatial anisotropy $\epsilon(b)$ which is the elliptic modulation of the final freeze-out hyper-surface at a given impact parameter vector\footnote{The impact parameter $\vec b$ is a vector which connects centers of the colliding nuclei in the plane perpendicular to the beam axis.} magnitude $b$, and by momentum anisotropy $\delta(b)$ which gives the modulation of the flow velocity profile. The events can be generated under several HYDJET++ model switches. The most realistic one, 'flow+quenched jets', includes both hydrodynamics expansion and quenched jets, and is used in the current analysis. The details of the model can be found in the HYDJET++ manual~\cite{Lokhtin:2008xi}.

\section{Results}
\label{sec:res}
\begin{figure}
  \includegraphics[width=0.60\textwidth]{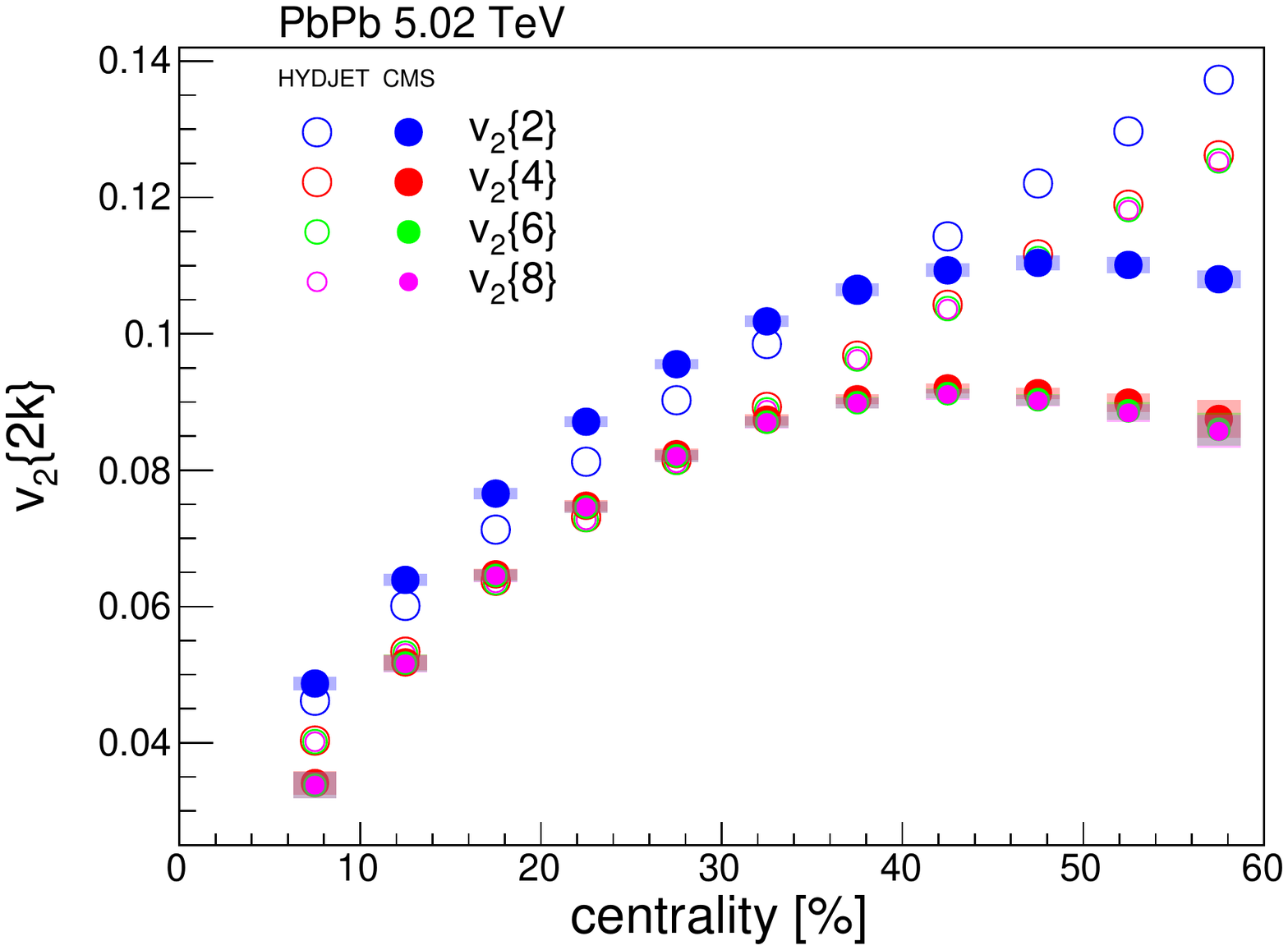}
\caption{\label{fig:1} (Color online) Elliptic flow harmonics of different cumulant orders $v_{2}\{2k\}$, ($k=1,...,4$) obtained from PbPb collisions at $\sqrt{s_{_{\mathrm{NN}}}}$ = 5.02~TeV generated by HYDJET++ model and using the experimental CMS data (taken from~\cite{Sirunyan:2017fts}) are shown with open and closed symbols as a function of the collision centrality. Data covers 0.3 $ < p_{T} < $ 3.0~GeV/c and $ |\eta| < $~1.0 range. The shadow boxes represent the systematic uncertainties of the experimental results, while the statistical uncertainties are smaller than the symbol size.}
\end{figure}

The centrality dependence of the elliptic flow harmonics obtained from different cumulant orders $v_{n}\{2k\}$ ($k=1,...,4$) extracted from PbPb collisions generated by the HYDJET++ model at 5.02~TeV incident energy are shown as open symbols in Fig.~\ref{fig:1}. In order to compare these results with the experimental ones, in the same figure are also shown corresponding CMS results taken from~\cite{Sirunyan:2017fts}. For each centrality interval, ranged from the central 5--10\% to rather peripheral 55--60\%, a cumulant analysis is performed within 0.3~$ < p_T < $~3.0~GeV/c and $|\eta| <$~1.0.

\noindent{Both, theoretical and experimental results exhibit a characteristic ordering between cumulants of different order: $v_{2}\{2\} > v_{2}\{4\} \approx v_{2}\{6\} \approx v_{2}\{8\}$ for all centralities. The difference between the $v_{2}\{2\}$ and higher order cumulants is more pronounced in the experimental data than in the HYDJET++ predictions. Qualitatively, HYDJET++ model properly predicts centrality dependence of the $v_{2}\{2k\}$. A relatively good agreement between the HYDJET++ predictions and the experimental data is achieved for more central collisions, while discrepancy between them becomes more pronounced going to more peripheral collisions.}

\begin{figure}
  \includegraphics[width=0.325\textwidth]{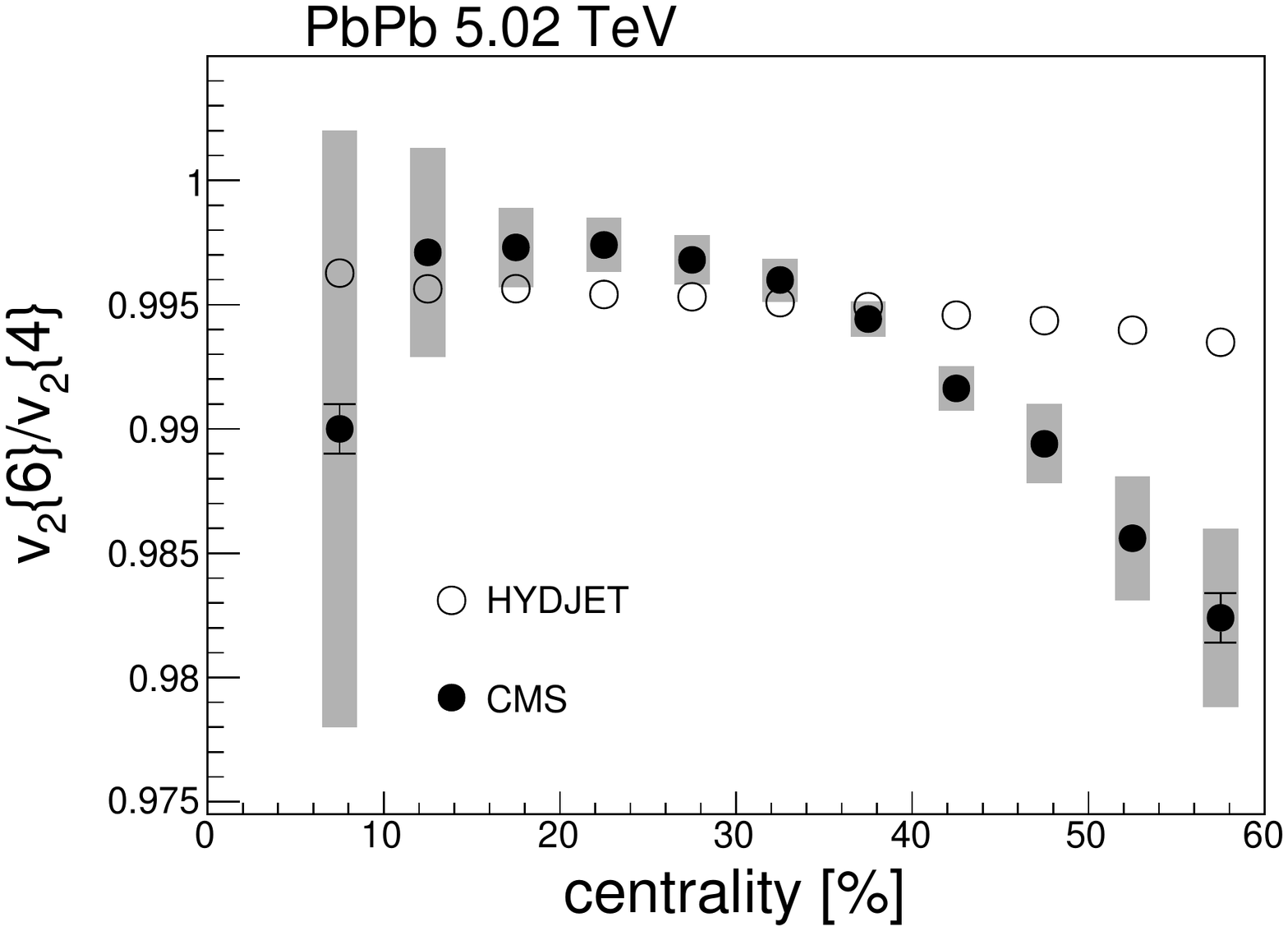}
  \includegraphics[width=0.325\textwidth]{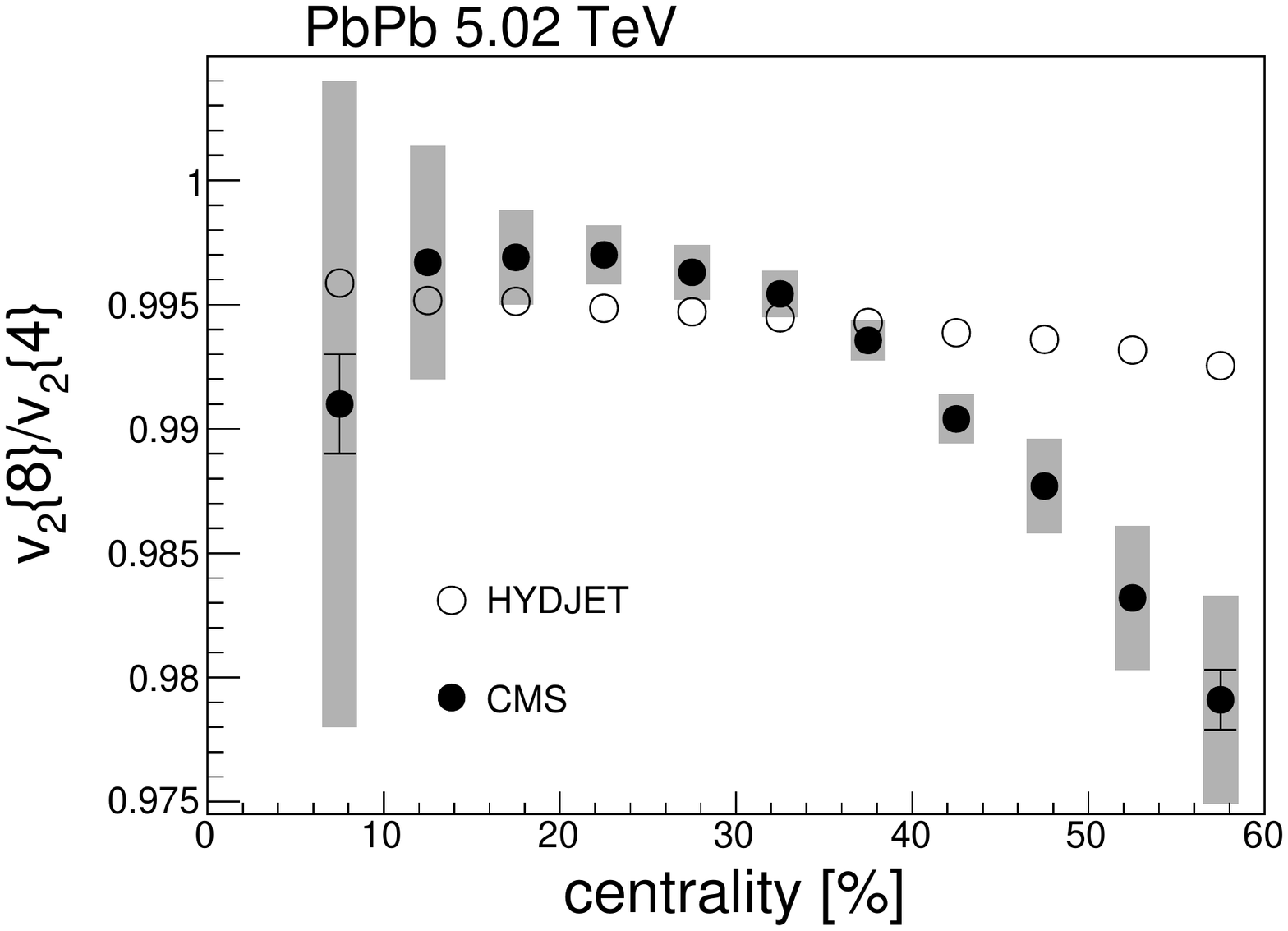}
  \includegraphics[width=0.325\textwidth]{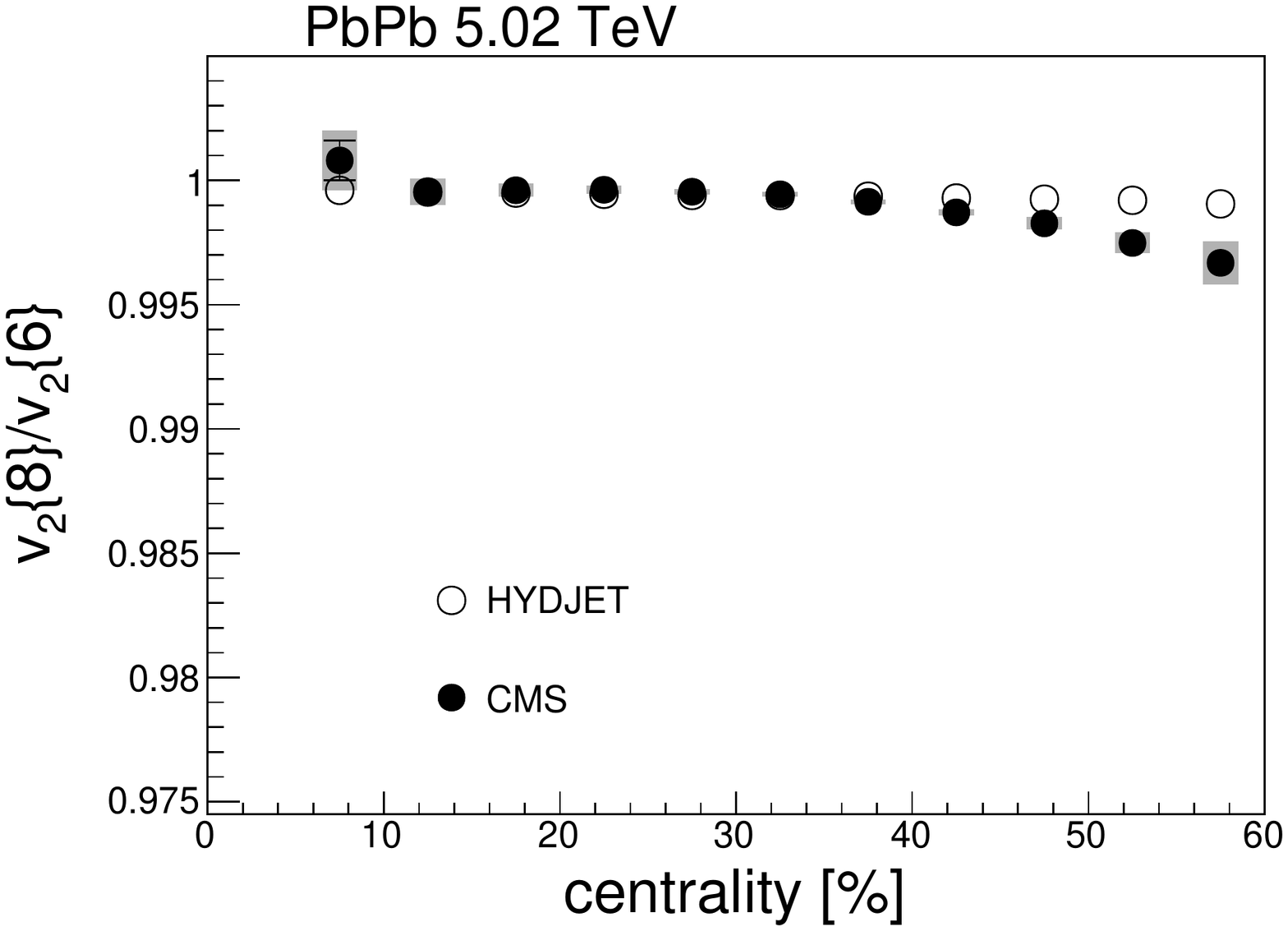}
  \caption{\label{fig:2} Centrality dependencies of the ratios of higher order elliptic flow cumulants: $v_{2}\{6\}/v_{2}\{4\}$ (left panel), $v_{2}\{8\}/v_{2}\{4\}$ (middle panel), and $v_{2}\{8\}/v_{2}\{6\}$ (right panel) obtained from PbPb collisions at $\sqrt{s_{_{\mathrm{NN}}}}$ = 5.02~TeV. The HYDJET++ model predictions are shown with open symbols, while the experimental CMS data (taken from~\cite{Sirunyan:2017fts}) are shown with closed symbols. Data covers 0.3 $ < p_{T} < $ 3.0~GeV/c and $ |\eta| < $~1.0 range. The error bars represent the statistical uncertainties. The shadow boxes represent the systematic uncertainties of the experimental results.}
\end{figure}

As from Fig.~\ref{fig:1} the rank ordering between the higher order cumulants is not visible well enough, in Fig.~\ref{fig:2} are shown centrality dependencies of ratios of the elliptic flow coefficients obtained for different cumulant orders. As for all centrality regions the ratios are smaller then one, they indicate a following rank ordering: $v_{2}\{4\} > v_{2}\{6\} > v_{2}\{8\}$. This confirms inconsistency with a pure Gaussian fluctuations model of the $v_{2}$ harmonics. The differences are smaller than 1\% and slightly increase going from central to peripheral collisions. In the same figure are shown CMS experimental results taken from~\cite{Sirunyan:2017fts}. In contrast to the HYDJET++ predictions, the experimentally measured ratios show much stronger centrality dependence and the deviation from unity reaches even a few percent in most peripheral events. A relatively good agreement between the experimental data and HYDJET++ predictions exists only for rather central events (up to 40\% centrality) and especially for the $v_{2}\{8\}/v_{2}\{6\}$ ratio.

\begin{figure}
\includegraphics[width=0.60\textwidth]{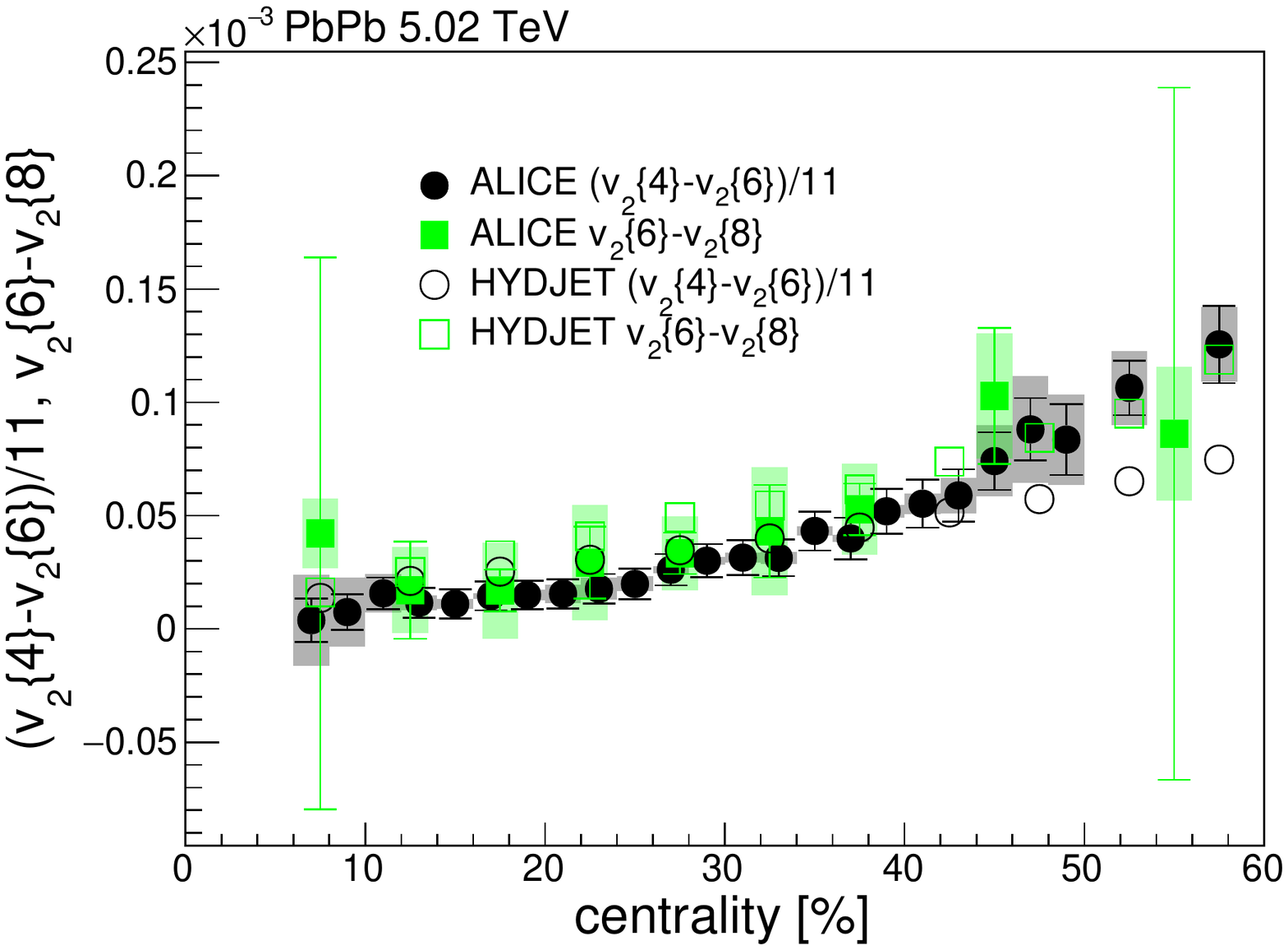}
\caption{\label{fig:3} (Color online) Centrality dependence of the differences of the $v_{2}$ Fourier harmonic calculated from different multi-particle cumulants extracted from PbPb collisions simulated with HYDJET++ model and from the ALICE~\cite{Acharya:2018lmh} experimental data at $\sqrt{s_{_{\mathrm{NN}}}}$ = 5.02~TeV. The error bars represent the statistical uncertainties. The shadow boxes represent the systematic uncertainties of the experimental results.}
\end{figure}

Figure~\ref{fig:3} shows $(v_{2}\{4\}-v_{2}\{6\})/11$ and $v_{2}\{6\}-v_{2}\{8\}$ quantities as a function of centrality in PbPb collisions at $\sqrt{s_{_{\mathrm{NN}}}}$ = 5.02~TeV simulated with the HYDJET++ model and measured by ALICE~\cite{Acharya:2018lmh}. The quantities extracted from the experimental data are observed to be in agreement, which demonstrates the validity of Eq.~\ref{UnEq}. The corresponding quantities extracted from the HYDJET++ simulation are observed to be not only in a mutual agreement but also in agreement with the experimentally measured ones.

\begin{figure}
  \includegraphics[width=0.60\textwidth]{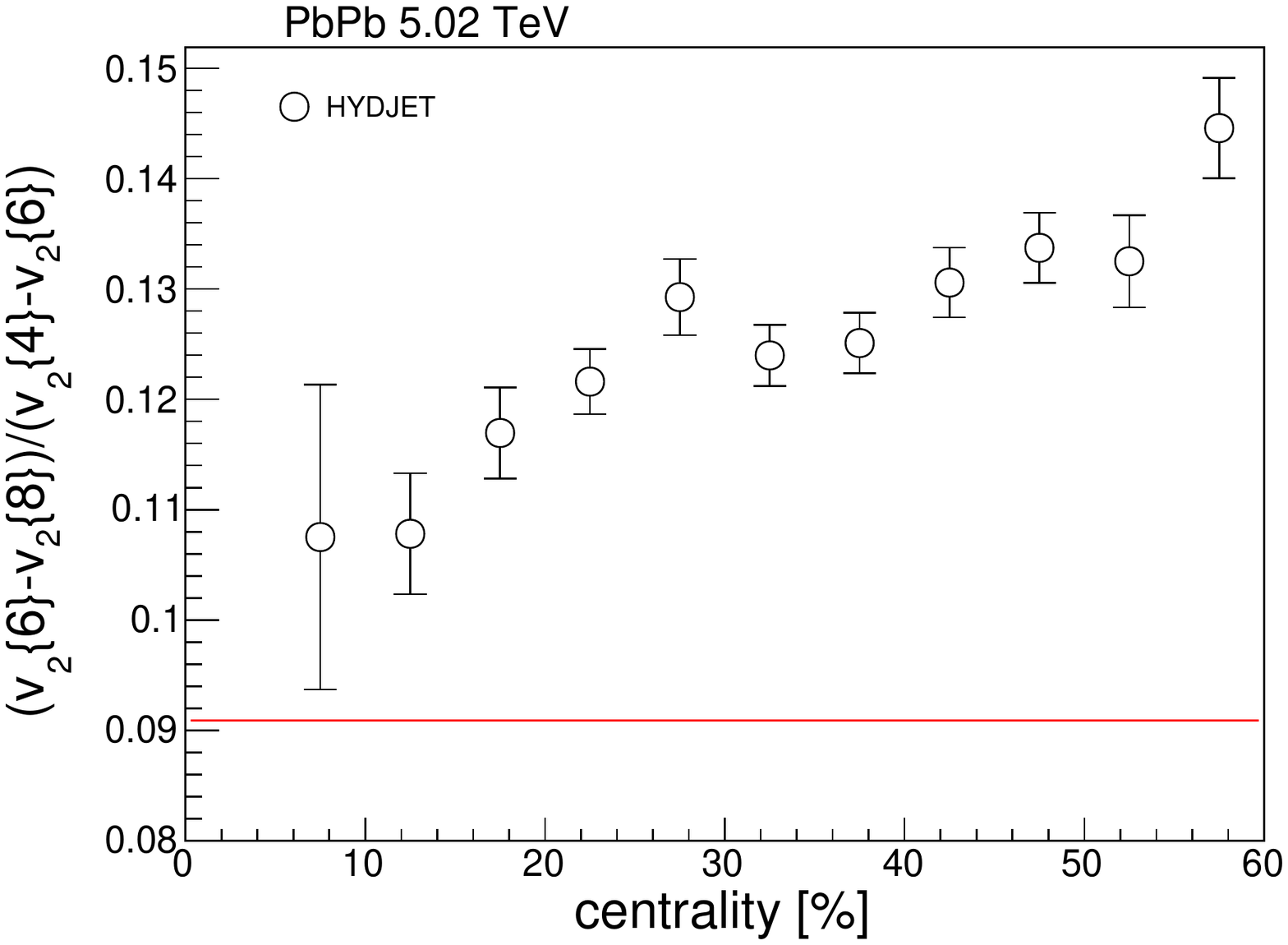}
  \caption{\label{fig:4} The centrality dependence of the ratio $(v_{2}\{6\}-v_{2}\{8\})/(v_{2}\{4\}-v_{2}\{6\})$ extracted from PbPb collisions simulated with HYDJET++ model at $\sqrt{s_{_{\mathrm{NN}}}}$ = 5.02~TeV. With the red horizontal line is indicated theoretical prediction of $\approx 0.091$. The analysis is performed for 0.3 $ < p_{T} < $ 3.0~GeV/c and $ |\eta| < $~1.0 range. The error bars represent the statistical uncertainties.}
\end{figure}

In order to check a hydrodynamic behavior of the medium simulated with the HYDJET++ model, in Fig.~\ref{fig:4} is plotted the ratio: $(v_{2}\{6\}-v_{2}\{8\})/(v_{2}\{4\}-v_{2}\{6\})$ for PbPb collisions at $\sqrt{s_{_{\mathrm{NN}}}}$ = 5.02~TeV. According to the Eq.~(\ref{UnEq}), in an ideal hydrodynamic behavior with a finite $v_{2}$ skewness one could expect that the plotted ratio should be equal to $\approx 0.091$. HYDJET++ model predicts an increase of that ratio going from central to peripheral collisions. It has the smallest and closest to the theoretical prediction value of about 0.11 at the most central 5--10\% analyzed collisions, and increases up to the value of 0.14 for the most peripheral collisions. The mean value of this ratio over the 5--60\% centrality range is 0.127~$\pm$~0.002, and is in an agreement with the experimental CMS and ALICE results of 0.18~$\pm$~0.08~\cite{CMSDurham} and 0.11~$\pm$~0.05~\cite{ALICEDurham} respectively\footnote{Statistical and systematical uncertainties in the experimentally measured ratio are too big to be plotted. It would be valuable to have better statistics in future data.}.

\begin{figure}
\includegraphics[width=0.60\textwidth]{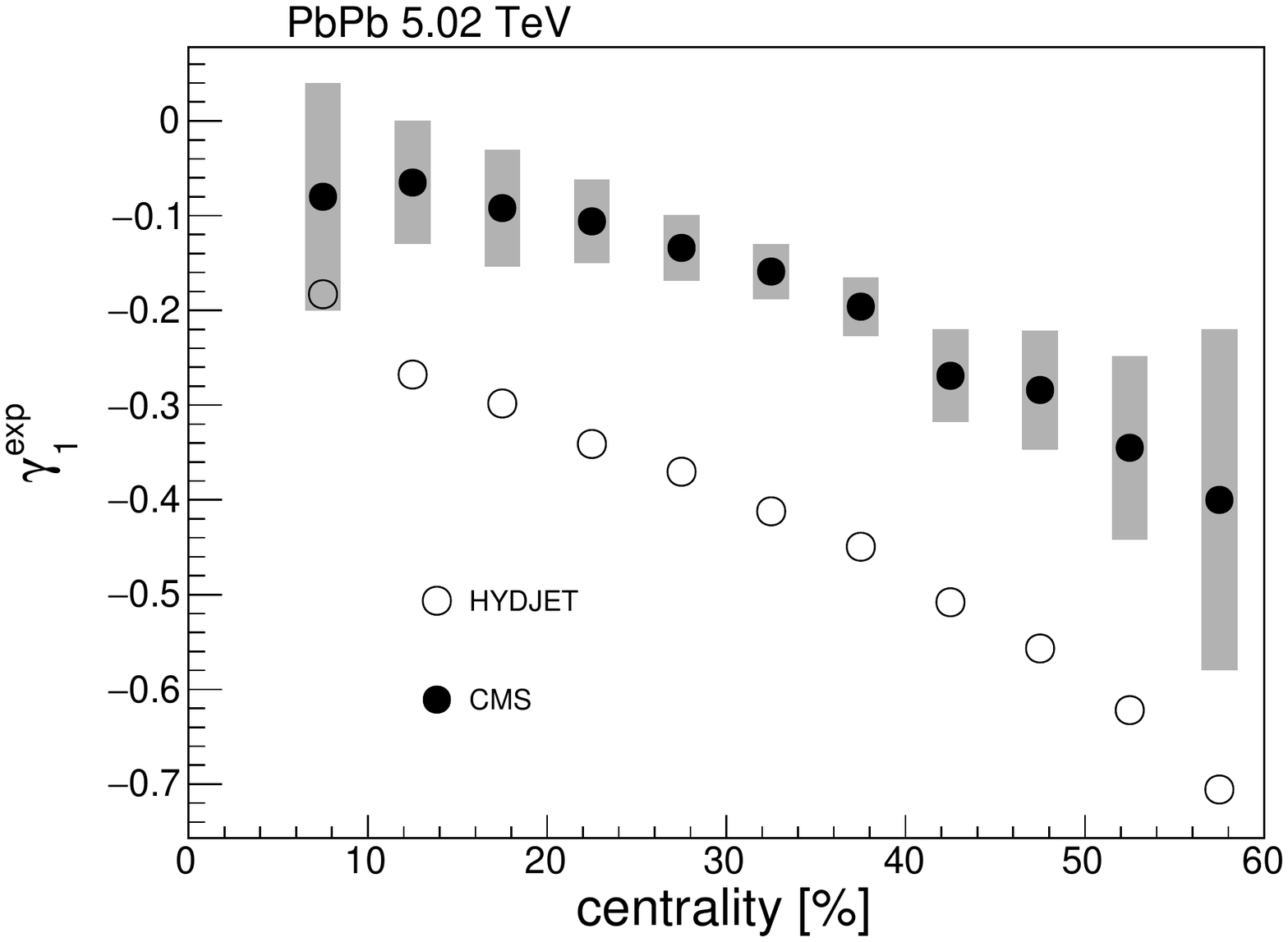}
\caption{\label{fig:5} The centrality dependence of the skewness calculated from the $v_{2}$ values of different cumulant orders in PbPb collisions at $\sqrt{s_{_{\mathrm{NN}}}}$ = 5.02~TeV. With open symbols is shown HYDJET++ model prediction, while  with closed symbols is shown the CMS experimental result (taken from~\cite{Sirunyan:2017fts}). The analysis is performed for 0.3 $ < p_{T} < $ 3.0~GeV/c and $ |\eta| < $~1.0 range. The shadow boxes represent the systematic uncertainties of the experimental results, while the statistical uncertainties are smaller than the symbol size.}
\end{figure}

Figure~\ref{fig:5} depicts centrality dependence of the elliptic flow skewness $\gamma^{exp}_{1}$ calculated using different cumulant orders in PbPb collisions at $\sqrt{s_{_{\mathrm{NN}}}}$ = 5.02~TeV. In the figure, both HYDJET++ model prediction and the experimental CMS results are obtained within the same 0.3 $ < p_{T} < $ 3.0~GeV/c and $ |\eta| < $~1.0 range. The CMS results are taken from~\cite{Sirunyan:2017fts}. Finite values for the skewness are obtained for both HYDJET++ model simulation and the experimental data. The shape of the $\gamma^{exp}_{1}$ distribution from the HYDJET++ model is qualitatively same as the one found from the experimental data, but the results from the model calculation and data differ significantly in their magnitudes. While the experimental measurements of $\gamma^{exp}_{1}$ is in a quantitative agreement with the theoretical predictions from~\cite{Giacalone:2016eyu}, the HYDJET++ model gives a much stronger $\gamma^{exp}_{1}$ deviation from zero. It is because the difference between the $v_{2}\{2\}$ and $v_{2}\{4\}$ in HYDJET++ model is significantly smaller wrt the same difference in the experimental data (see Eq.(\ref{Skew})). The CMS results are in an agreement with the ALICE experimental results~\cite{Acharya:2018lmh} too.

\section{Conclusions}
\label{sec:conc}
The cumulant analysis method for the $v_{2}$ elliptic flow coefficients in PbPb collisions generated by the HYDJET++ model at $\sqrt{s_{_{\mathrm{NN}}}}$ = 5.02~TeV shows that the event-by-event fluctuations in the $v_{2}$ magnitude are not Gaussian. The analysis is performed as a function of centrality covering the range from 5\% up to 60\% collision centralities. As expected the $v_{2}\{2\}$ clearly has a magnitude larger than the ones from the higher order cumulant. But, a rank ordering between the higher order cumulants: $v_{2}\{4\} > v_{2}\{6\} > v_{2}\{8\}$, with differences smaller than one percent, is also observed. Comparison of the $(v_{2}\{4\}-v_{2}\{6\})/11$ and $v_{2}\{6\}-v_{2}\{8\}$ distribution shows that the HYDJET++ predictions are in a good agreement with the ALICE data~\cite{Acharya:2018lmh}. A hydrodynamic check for the centrality dependence of the ratio $(v_{2}\{6\}-v_{2}\{8\})/(v_{2}\{4\}-v_{2}\{6\})$ shows that HYDJET++ model gives an increasing distribution with the mean value closed to the expectation from the ideal hydrodynamics similarly as is observed in the experimental CMS~\cite{CMSDurham} and ALICE~\cite{ALICEDurham} data. In the case where there is a difference in the magnitudes from the higher order cumulant, the standardized skewness $\gamma^{exp}_{1}$ is found to be negative with an increasing magnitude as collisions become less central. The HYDJET++ model qualitatively predicts correct behavior of the skewness centrality dependence, but gives significantly larger magnitude of the $\gamma^{exp}_{1}$ than the experimental result.

\begin{acknowledgments}
The authors acknowledge the support of the Bilateral Cooperation 
between Republic of Serbia and People's Republic of China 451-03-478/2018-09/04 ''Phenomenology in high energy physics''. 
\end{acknowledgments}

\end{document}